  \documentclass{cambridge6A}
  \usepackage{natbib}

  \usepackage{rotating}
  \usepackage{floatpag}
  \rotfloatpagestyle{empty}

  \usepackage{amsthm}
  \usepackage{graphicx}

  \usepackage{multind}
  \makeindex{authors}
  \makeindex{subject}

  \theoremstyle{plain}

  \theoremstyle{definition}

  \theoremstyle{remark}

\begin{document}

  \title[Asteroseismology observations and space missions]
    {Extraterrestrial seismology}



  \frontmatter
  \maketitle
  \tableofcontents
  \listoffigures

  \mainmatter

\chapter{Asteroseismology observations and space missions}
\vspace{-3cm}
\begin{center}
{\bf T.Appourchaux}\\
Institut d'Astrophysique Spatiale, UMR8617, Universit\'e Paris-Sud, B\^atiment 121, 91405 Orsay Cedex, France\\
\vspace{1cm}
{\bf F.Grundahl}\\
Stellar Astrophysics Centre, Department of Physics and Astronomy, Aarhus University, Ny Munkegade 120, Aarhus C, Denmark
\end{center}
\vspace{2cm}
\section{Introduction}
Variable stars have been observed since the discovery of the variability of Mira by David Fabricius at the end of the XVI$^{\rm th}$ century \citep{Olbers1850}.
The variation in luminosity of Mira is 1700 times its basic flux or variation of 10 magnitudes !  The variations are large enough for not having to use reference stars.
The first known photometric measurement of variability was done by \citet{Goodricke1783} for Algol using stars differing by two magnitudes only.  Using the same comparison technique, \citet{Goodricke1785} discovered that $\delta$ Cepheid was a variable star of a peculiar type showing a periodical non-sinusoidal variation of about one magnitude.  As a matter of fact the discovery of the first $\delta$ Cepheid star was indeed performed a few month before by \citet{Pigott1785}.

With the invention of photography by Nic\'ephore Ni\'epce in 1826, the detection of variable star was soon to become less driven by visual assessment.  It is more than half a century later that \citet{Roberts1889} suggested the use of photography for studying variable star.  The technique for detecting variable stars was quickly improved by recording photographically variations in the stellar spectra \citep{Fleming1895} which would lead to visual identification of these variables \citet{Reed1893}.  

The introduction 
of selenium photometer by \citet{Stebbins1907} introduced the electric determination of variability of stars \citep{Stebbins1911}.  Finally, the development of the photo-electric cell as anticipated by \citet{Stebbins1915} was the start of an era which has not ended today: the measurement of photon quanta using the photoelectric effect.  The improvement in the detectability of variable stars lead to the discovery of Rapidly Oscillating Ap Stars by \citet{Kurtz1978} oscillating with periodicity of about 6 to 10 minutes with an amplitude of about a thousandth of a magnitude (mmag).  \citet{Kurtz1986} were able to reach a noise limit of about 0.07 mmag on HR3831, and even down to 0.02 mmag with longer data sets \citep{Kurtz2000}.  

From space, lower levels of variabilities had already been detected  on the Sun using the ACRIM\footnote{Active Cavity Radiometer Irradiance Monitor} instrument aboard the Solar Maximum Mission \citep{Woodard1983}; the noise level was about one part-per-million (close to one $\mu$mag).  These measurements on the Sun were reproduced from ground-based observatories with various levels of successes \citep{Schmidt1984,Jimenez1987}.  Clear detection of p modes in intensity from the ground was made possible using the property of the atmosphere combined to the significant angular diameter of the Sun \citep{TA95a}.  Unfortunately due to the minute angular diameter of the stars, the intensity fluctuations are not limited by transparency fluctuations but by scintillation \citep{Young1967}.

The observation of variable stars possesses some astrophysical importance for the understanding of the physics in our Universe.  The variety of variability observed since the XVI$^{\rm th}$ century lead to the discovery of various classes of variable stars:
\begin{itemize}
\item Cepheids, RR Lyrae, $\delta$ Scuti, Mira type : stars having their variable luminosity scaled to their mass
\item $\beta$ Cephei : stars pulsating due to the opacity change mechanism ($\kappa$ mechanism)
\item $\gamma$ Doradus : stars having gravity modes excited by convection \citep{Dupret2005}
\item Rapidly oscillating Ap stars : having their pulsation axis aligned with the axis of their magnetic field 
\item White dwarves, solar-like oscillations :  stars oscillating non radially and  having up to million of modes excited
\end{itemize}
This list of variable stars is a sub-set of what is available across the Hertzprung-Russell diagram \citep{Eyer2008}.  

\begin{figure}
\center{
\hspace{0.75cm}\includegraphics[width=0.7\textwidth,angle=0]{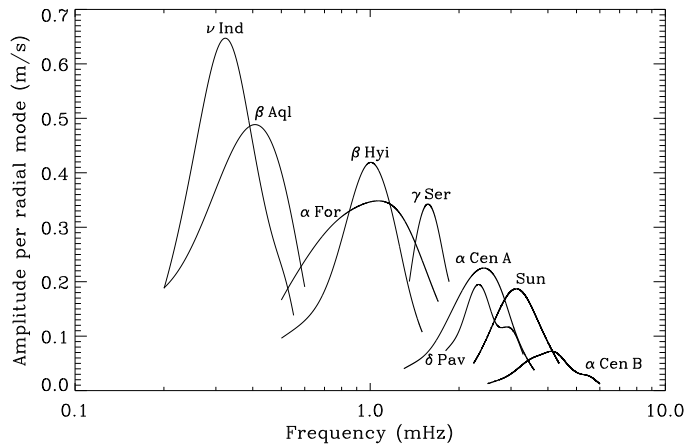}
}
\caption[Envelope of solar-like oscillation observed in stellar radial velocity]{Mode-amplitude envelope of solar-like oscillation observed in stellar radial velocity as a function of frequency for various stars \citep[Stolen from][]{Kjeldsen2008}.}
\label{amplitude}
\end{figure}

With so many variable stars, the detection of such periodicities lead to the development of a new type of science: asteroseismology.  This new field of Science was so baptised by \citet{JCD1984a} as {\it "The Science of using stellar oscillations for the study of the properties of stars, including their internal structure and dynamics"}.  In this seminal paper, \citet{JCD1984a} introduced what could be done on stars similar to the Sun.  The feasibility of detecting stellar oscillations similar to those of the Sun was theoretically given by \citet{CD83}.  They provided amplitude of about 10\,cm.s$^{-1}$ for radial velocity and few tens of ppm for intensity fluctuations.  The detection of the first solar-like p modes by in another star by \citet{FB2002} was the real of asteroseismology.   A few years later, several other stars were already exhibiting solar-like p modes as shown in Figure~\ref{amplitude}.  This what the start of a new era for stellar physics.  

The goal of this paper is to provide a review of how one can detect and observe such solar-like oscillations.  The first section will focus on the scientific requirements derived from the properties of the oscillations.  The second section will treat of the history of the first detection attempts that eventually lead to the implementation of more ambitious projects described in the third section.  Finally with the last section, we conclude with a focus on the next generation of ground- and space-based projects.

\section{What are the requirements}
\subsection{How to recover the signal?}
The design of the instrumentation capable to detect solar-like oscillations will be driven by three factors:
\begin{itemize}
\item the highest frequencies of the modes
\item the mode amplitude (in radial velocity of intensity)
\item the mode lifetime 
\end{itemize}
The first number will provide the sampling time of the time series.  The second and third will provide the level to which the instrument should be stable and the duration of the observations, respectively.

{\bf\noindent Sampling time.}  As for the first number, this is closely related to the acoustic cut-off of the frequencies.  Since most of the excited modes in solar-like stars are pressure modes
there is a frequency at which the modes are not reflected anymore at the surface of the star (i.e. there are no eigenmodes anymore).  The acoustic cut-off $\nu_{\rm ac}$ was given by \citet{Brown1991} as:
\begin{equation}
\nu_{\rm ac} \propto \frac{M}{R^2 \sqrt{T_{\rm eff}}}
\label{freq_ac}
\end{equation}
where $M$, $R$ and $T_{\rm eff}$ are the stellar mass and radius in units of the solar one, and the effective temperature.  The highest cut-off frequency is obtained for stars smaller than the Sun having a larger surface gravity.  For stars with mass of 0.8 $M_{\odot}$ and luminosity of 0.4 $L_{\odot}$, the acoustic cut-off frequency scaled to that of the Sun is about 10200 $\mu$Hz.  The application of Shannon's theorem implies that the maximum sampling time should be less than 49 s ($\approx$1/2/0.0102).  In order to minimise the high frequency noise, it is also advisable to integrate the signal over the sampling time, i.e. to have a duty cycle close 100\%. \citep[See][for an application of various digital-world approaches to asteroseismology]{Appourchaux2011}.


{\bf\noindent Background noise.}  The mode amplitude was theoretically derived by \citet{CD83} and \citet{GH99}.  The latter theoretically confirmed the scaling law provided by \cite{HK1995}:
\begin{equation}
\frac{A}{A_{\odot}} \approx \frac{L}{M}
\label{amp}
\end{equation}
where $A$ is the maximum stellar mode amplitude, $A_{\odot}$ is the maximum solar mode amplitude and $L$ is the stellar luminosity in unit of the solar one.
The units of $A$ are either cm.s$^{-1}$ or ppm.  The maximum of the mode amplitude is obtained for a frequency $\nu_{\rm max}$ which follows the same scaling relation as Eq.~(\ref{freq_ac}) \citep{Brown1991}.  Since the stellar luminosity scales like $M^4$, the lowest mode amplitude is obtained for star lighter than the Sun.  The  maximum mode height in the spectral density is directly linked to the maximum mode amplitude and the mode linewidth as:
\begin{equation}
H=\frac{2 A^2}{\pi \Gamma}
\label{mh}
\end{equation}
where $H$ is the maximum mode height and $\Gamma$ is the mode linewidth.  The relation is derived assuming that the mode profile is Lorentzian.  Combining Eqs.~(\ref{freq_ac}), (\ref{amp}) and (\ref{mh}), one can obtain a relation between the mode height and the frequency of maximum power as:
\begin{equation}
H=\frac{2T_{\rm eff}}{\pi \Gamma_{\rm max}} \frac{1}{\nu_{\rm max}^2}
\label{mh_th}
\end{equation}
The theoretical mode linewidths have been derived by \cite{GH99}.  They typically gave mode linewidth of about 2 $\mu$Hz at the location of the maximum power for solar-like stars.


\begin{figure}[!]
\center{
\hspace{0.75cm}\includegraphics[width=0.5\textwidth,angle=90]{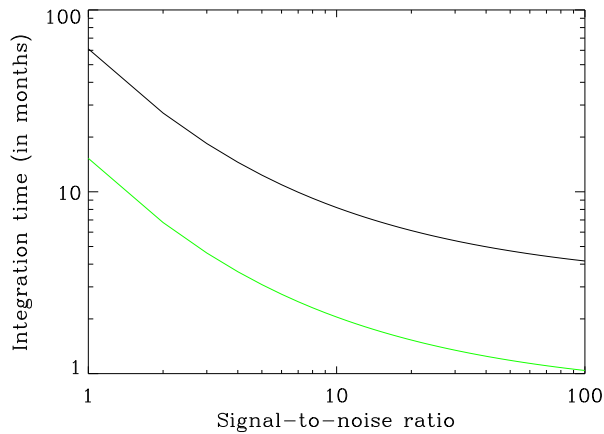}
}
\caption[Integration time as a function of signal-to-noise ratio]{Integration time as a function of signal-to-noise ratio in the power spectrum for having a mode frequency precision of 0.1 $\mu$Hz (Black line) and of 0.2 $\mu$Hz (Green line) for a 1-$\mu$Hz linewidth.  The mode is a singlet or an $l=0$ mode.}

\label{s-n-time}
\end{figure}

\begin{figure}[!]
\center{
\includegraphics[width=0.5\textwidth,angle=90]{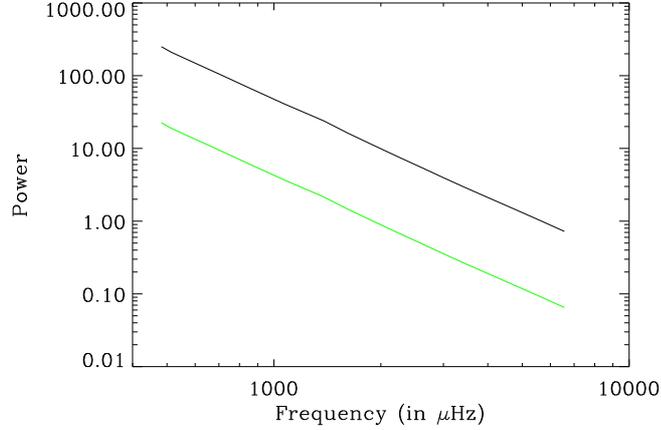}
}
\caption[Background noise as a function of the frequency of the maximum mode power]{Background noise as a function of the frequency of the maximum mode power required for detecting stellar p modes with a signal-to-noise ratio of 10, for stellar radial velocity (black), for intensity (green).  The typical mode linewidth is 2 $\mu$Hz.  For stellar radial velocity the unit are (cm/s)$^{2}\mu{\rm Hz}^{-1}$, while for intensity they are ppm$^2\mu{\rm Hz}^{-1}$.  The highest frequency corresponds to 0.8 $M_{\odot}$ star, while the lowest frequency corresponds to 2 $M_{\odot}$ star.}
\label{noise_level}
\end{figure}

In order to detect the modes at the maximum of power, it is reasonable to put an upper limit to the background noise of three times lower (target) and 10 times lower (goal) than the mode height given by Eq.~(\ref{mh_th}).  The background noise is composed of several source noise: the instrumental noise that can be controlled and reduced and the stellar noise that cannot be controlled.  The goal of a signal-to-noise ratio of 10 is derived from the theoretical expression of the mode frequency precision provided by \citet{KL92} as:
\begin{equation}
\sigma_{\nu}=\sqrt{f(\beta) \frac{\Gamma}{4\pi T}}
\end{equation}
where $\beta$ is the inverse of the signal-to-noise ratio (=$H/B$ where $B$ is the background noise), $\Gamma$ is the linewidth of the mode profile assumed to be Lorentzian and $T$ is the observation time, and $f(\beta)=\sqrt{1+\beta}(\sqrt{1+\beta}+\sqrt{\beta})^3$.  With a signal-to-noise ratio better than 10, a frequency resolution smaller than 0.2 $\mu$Hz can be obtained for integration time longer than 2 months, and for signal-to-noise ratio better than 3 for time longer than 4 months (See Figure~\ref{s-n-time}).  These signal-to-noise ratio figures of 3 and 10 are typical upper limit that can be tweaked according to various engineering and scientific strategies.  The tweaking can certainly be done for having even lower noise values.  Figure~\ref{noise_level} provides the value of the background noise (goal) for stellar radial velocity and intensity, assuming that $A_{\odot}$ is about 10\,cm/s and 3\,ppm for the Sun.  

For the lightest star of 0.8 $M_{\odot}$, the values are 0.8 (cm/s)$^{2}\mu{\rm Hz}^{-1}$ and 0.07\,ppm$^{2}\mu{\rm Hz}^{-1}$.  Assuming that the noise power spectrum is white from frequency 0 to the Nyquist frequency, we can deduce that the maximum rms noise is 1.1\,m/s and 32\,ppm in radial velocity and intensity, respectively.  For the heaviest star, the corresponding levels are 17.4\,m/s and 505\,ppm.  If one assumes a $1/\nu$ noise starting at some frequency lower than the lowest mode frequency, then these levels can be somewhat increased.

{\bf\noindent Observation duration.}  Finally, the duration of the observation will be driven by the mode linewidth or their lifetime.  The mode lifetime is related to the mode linewidth by the following relation:
\begin{equation}
\tau=\frac{1}{\pi \Gamma}
\end{equation} 
In order to resolve the mode linewidth, the observation time should be such as to have a resolution three times better than this linewidth which implies using the previous equation that the observation time is given by:
\begin{equation}
T=3\pi \tau
\end{equation}
It means that we need to observed 10 lifetimes of the mode for resolving them.  For lifetime of 3.7 days (=1 $\mu$Hz), the observation duration should be at least 37 days.  The modes at the low frequency end have long lifetimes \citep{GH99} such that observation of typically 1 year long would give access to linewidth of 0.1 $\mu$Hz.

\subsection{How can we measure?}  
Unlike the Sun for which we can see its surface, the stars are far reaching.  Their angular diameter ranges from about 10 $\mu$arcsec to a maximum of 6 arcsec for $\eta$ Carinae \citep[See the catalog of][]{PF2001}, and about 30 stars have an angular diameter known to be larger than 100 milliarcsec (mas).  It is clear that the imaging of the surface of such stars with about 1000 resolution elements would require telescope of diameter larger than 1000 m in diameter.  Therefore, it is much easier to measure the stellar radial velocities or intensity without any resolution, i.e. integrated over the stellar disk.  The effect of the integration over the stellar disk is known to filter out all modes above $l=4$ in velocity \citep{Dziembowski77,C-DG82}, and above $l=3$ in in intensity \citep{Dziembowski77,Toutain1993}.

The measurement of radial velocity for detecting p modes was performed on the Sun using resonance cell \citep{AC79,Grec83}.  For intensity measurement, irradiance measurements lead to the detection of p modes using the ACRIM radiometer \citep{Woodard1983}.  For the stars, these two techniques needed to be adapted for coping with the  lack of photons.  

\begin{figure}
\center{
\includegraphics[width=0.6\textwidth,angle=90]{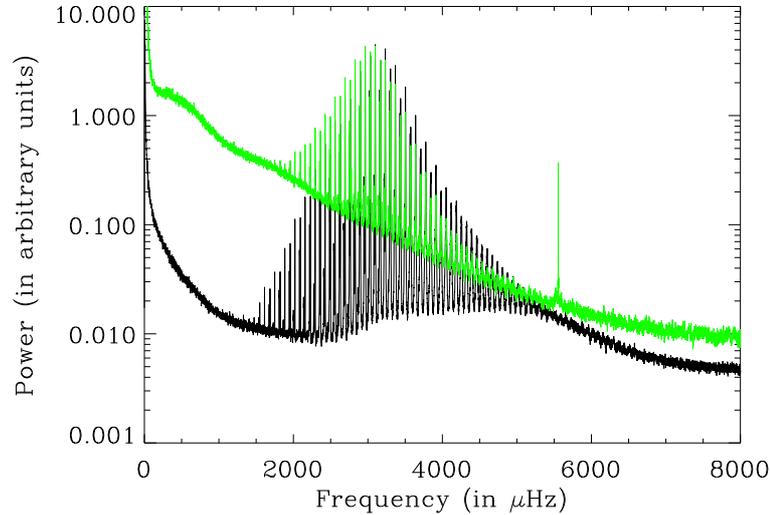}
}
\caption[Solar radial velocity vs intensity power spectra]{Power spectra as a function of frequency for the Sun smoothed over 1 $\mu$Hz: solar radial velocity measured by GOLF (black line), and intensity measured by the LOI (Green line).  The GOLF time series is 12.75 year long while the LOI is 16.9 year long.  Power units are arbitrary.  The large spike at 5555 $\mu$Hz is an artefact of the basic cadence of data acquisition system of VIRGO which is 3 minutes.}

\label{golf_loi}
\end{figure}

The main differences between stellar radial velocity and intensity measurements are summarised in Fig.~\ref{golf_loi} which is a comparison of the results obtained looking at the Sun as a star with the GOLF\footnote{Global Oscillations at Low Frequency} instrument \citep{AG95} and of the LOI\footnote{Luminosity Oscillations Imager} of the VIRGO\footnote{Variability of Irradiance and Gravity Oscillations} instrument \citep{TABA97,CFJR95}; both instruments are aboard the SoHO\footnote{Solar and Heliospheric Observatory} spacecraft.  The essential difference lies in the effect of granulation which is much smaller in solar radial velocity compared to intensity.  As a result the maximum signal-to-noise ratio is about 30 in intensity and as high as 300 for solar radial velocity.  The added benefit of radial velocity vs intensity measurements is then to be able to detect more modes (typically twice as many) and to detect modes with longer lifetimes up to a year \citep{Salabert2009}.

\vspace{0.5cm}
{\bf \noindent Stellar radial velocity.}
The measurement of stellar radial velocity on single absorption line is challenging because of the lack of photons.  The difficulty in the design of such spectrophotometer is to make an image of the stellar spectrum over a wide wavelength band without loosing photons and at high spectral resolution.  Spectrophotometer such as the HARPS\footnote{High Accuracy Radial velocity Planetary Search} instrument is designed for high spectral stability using a spectral calibration lamp for wavelength reference, high spectral resolution (R=120000), high efficiency and vacuum operation \citep{Pepe2000}.  The HARPS instrument was the evolution of several forefathers such as CORAVEL \citep{Baranne1977,Baranne1979}, ELODIE \citep{Baranne1996} and CORALIE \citep{Queloz2000}.

The seminal work that lead to the development of spectrometer able to detect very small radial velocities is due to \citet{Connes1985}.  In this paper, he could predict what would be the limit provided by a 1-m diameter telescope on a 5th magnitude star similar to the Sun; the limit was 0.36 (cm/s)$^{2}\mu{\rm Hz}^{-1}$ which roughly corresponds to using 3000 absorption lines instead of one line \citep{Harvey1988}.  The working principle of the spectrometer of \citet{Connes1985} was to use an echelle spectrograph  for measuring the displacement of many absorption lines, coupled with a low finesse Fabry-Perot and stabilised laser.  The measurement of the radial velocity was not done using correlation but using an optimal weighting scheme.  Given the shear complexity of the spectrophotometer, only one star can be studied at any one time; making the measurement a mono-object technique.  The precision reached by such a technique depends primarily on: 
\begin{itemize}
\item the number of photons gathered (telescope diameter, optic efficiency)
\item the spectral coverage
\item the resolving power of spectrometer 
\item the star rotational velocity
\end{itemize}
\citet{BouchyQueloz} studied the impact of all these parameters on the velocity noise induces by photon noise.  They concluded that a spectrometer with a spectral resolution better than 5 10$^{4}$, with a spectral coverage of 300 nm and a telescope diameter of a one meter would give a velocity spectral density induced by photon noise of typically 0.9 (cm/s)$^2\mu{\rm Hz}^{-1}$ for a slowly rotating star like the Sun, sufficient for detecting stellar p modes according to Figure~\ref{noise_level}.

The concept described in \citet{Connes1985} was implemented in the EMILIE spectrograph \citep{Schmitt2000} and then used in a simplified version by his former student for detecting p modes on $\alpha$ Cen \citep{Bouchy2001} with the CORALIE instrument.

\begin{figure}
\center{
\includegraphics[width=0.6\textwidth,angle=90]{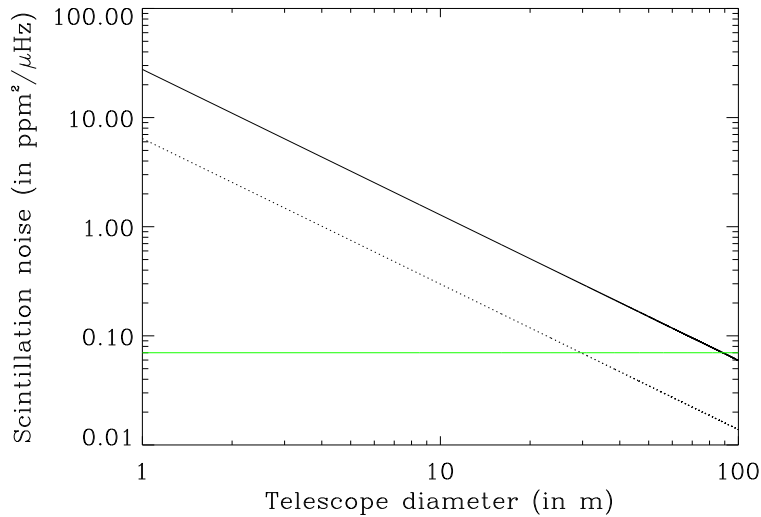}
}
\caption[Scintillation noise as a function of telescope diameter ]{Scintillation noise as a function of telescope diameter as given by Eq.~\ref{young} for an airmass of $X=1.5$ (solid line) and of $X=1.0$ (dotted line) for an observatory at an altitude of 4000 m.  The green line is the goal sets as 0.07\,ppm$^2\mu{\rm Hz}^{-1}$ for 0.8 $M_{\odot}$ star.}

\label{scinti}
\end{figure}

\vspace{0.5cm}
{\bf \noindent Stellar intensity.}
In contrast to stellar radial velocity, the main advantage of doing photometry with CCD arrays is the possibility to observe many objects at any one time not only one object.  
Several tentatives were done by \citet{Gilliland1988, Gilliland1991, Gilliland1992, Gilliland1993} for doing asteroseismology from the ground using such arrays.   These  trials were not successful because the measurement of stellar intensity from the ground is limited mainly by the effect of scintillation not by photon noise.  The angular diameter of stars being typically of the order of few marcsec which combined with a seeing of 0.1 to 0.5 arcsec produces large intensity fluctuations as shown by \citet{Young1967} as:
\begin{equation}
\sigma=0.09 D^{-2/3} X^{1.8} e^{-h/8000} 
\label{young}
\end{equation}
where $\sigma$ is the spectral density of the scintillation noise in $1/\sqrt{{\rm Hz}}$, $D$ is the telescope diameter in cm, $X$ is the airmass at the observatory and $h$ is the observatory altitude in meter, an example of the application of the formula is shown on Fig.~\ref{scinti}.   It was already clear back in the 1990's that telescope diameter as large as 50 m was not feasible option for beating down the scintillation noise \citep{PRISMA91,PRISMA93}.  The alternative developed at the times were already to do space-based observations instead of ground-based observations. 

For space-based observations, the main source of noise besides photon noise is the combined effect of pointing jitter and of CCD pixel-to-pixel response non-uniformity (PRNU).  The fact that the stars are nearly punctual imposes to reduce the effect of pointing jitter by defocusing the image of the stars on the detector.  \citet{Gruneisen1998} provided the photometric noise induced by pointing jitter and by the as:
\begin{equation}
\sigma_{\rm jitter}=\sqrt{2} \frac{{\rm PRNU} \delta x}{N^{3/4}}
\end{equation}
where $\sigma_{\rm jitter}$ is the relative photometric variation, $\delta x$ is the jitter in units of CCD pixels, PRNU is in relative unit and $N$ is the equivalent number of pixels making the defocused image of the star.  For a typical PRNU of 1\%, a pointing jitter of 0.1 pixel rms and a typical number of pixel of 300, the pointing jitter is about 20\,ppm which imposes typical pointing stability better than 1 arcsec peak-to-peak.  Such levels have been obtained with the CoROT\footnote{Convection and Rotation of Stars} mission using the science detector as a stellar attitude sensor, and pointing jitter corrections were even devised for compensating larger pointing effects \citep{Drummond2006,Fabio2007}.  As we will see in Section \ref{revolution}, the major source of noise for intensity measurements when the jitter noise is under control is solely the photon noise.


\vspace{0.5cm}
{\bf\noindent Ground vs Space} Most observations performed in the visible can be done from the ground.  The justification for going to space is usually based upon wavelength being blocked by the atmosphere (UV, EUV, X rays), the fullness of the sky coverage, the absence of day/night interruptions, thermal requirements (operations close to 0 K).  As for stellar radial velocity and intensity measurement, the sky coverage and the absence of interruptions would be enough to justify having a spectro / photometer puts in space.  In addition, we noted above that scintillation noise can be beaten by having very large telescopes that are not feasible. 

The day/night interruptions generate daily alias located at harmonics of 11.57 $\mu$Hz making the mode identification difficult when the mode separation is about that value.  It has been often argued that the presence of such interruptions could be overcome by having a network of telescope.  The debate gave rise in the 1990's to two type of projects in helioseismology:
\begin{itemize}
\item Ground-based network: the Global Oscillation Network Group (GONG) \citep{Harvey1996}; Birmingham Solar Oscillation Network (BiSON) \citep{YE91}; International Research on the Inside of the Sun (IRIS) \citep{Fossat1991}
\item Space mission: SoHO \citep{Domingo1995}
\end{itemize}
The canonical number of stations of 6 provides a data fill (in the case of the Sun) of nearly 90\% as theorised by \citet{Hill85} and as measured with the GONG network by \citet{Harvey1996}, but as low as 78\% for the BiSON network \citep{WJC1996}.  The typical reduction in the amplitude of the first alias is about $(1-D)^2$, where $D$ is the data fill, i.e. for a 90\% data fill, the alias amplitude is 1\% of the main peak, and for a 75\% data fill, the amplitude is 6\% of the main peak.  Since the development of network, techniques have been devised for compensating and correcting the impact of the daily aliases on the measurement of stellar p mode frequencies \citep{Stahn2008}.  

In theory, space missions provide better data fill if the interruptions are not periodic.  This is the case of the SOHO mission which has a data fill ranging from 90\% to 98\% but with non-periodic random-like behaviour.  This was not the case of the CoROT mission because the orbit period of nearly 100 minutes would provide periodic interruption / modulation by passing through the South Atlantic Anomaly; the resulting data fill was 90\% \citep{Appourchaux2008}. 

The advantage of space mission vs ground observation is mainly in the sky coverage that can be achieved for long duration without any interruption.  For instance, \citet{Mosser2007} showed that since the local time shifts by 4 min per day with respect to the sidereal time, the duration of observation of the stars cannot be longer than 5 months.
This applies to both stellar radial velocity and intensity measurements.  In addition, the number of stations needs to be increased since the observatories can observe either the boreal or the austral sky, i.e. stations are located in the Northern and Southern hemisphere.  The minimum number of stations of the Stellar Oscillation Network Group (SONG) is therefore 8 with 4 stations on either side of the Equator \citep{Grundahl2009, Grundahl2011}.  Space missions such as CoROT and {\it Kepler} have other limitations in terms of sky coverage and observing time.  Their limitations of sky accesses are mainly related to the availability of solar power in Earth centred orbits (solar panel drift, eclipses) providing typical observation duration of 6 months.  The sky coverage and observing time are greatly enhanced when the spacecraft is not orbiting the Earth.

In summary, what are the real advantages / disadvantages of space observation vs ground-based observations?  

The argument that is often used for preferring ground vs space is that it is less expensive.  We believe this is a rather naive argument because for the man in the street, doing Science is already expensive, and funding should be allocated to other type of development such as health care, medical care or environment preservation.  We could also argue that the maintenance of a network requires also funding.  Equipment replacement, damages due severe weather, field trip and so forth also need to be included in the required fund.  In comparison, space instruments are by design and construction more robust and reliable: there is no need for maintenance apart from the spacecraft station keeping.  A counter example is the correction of the short-sighted Hubble Space Telescope (HST) which needed to be corrected and was corrected thanks to a HST servicing mission.  This is the example of space mission designed with a ground-based type of approach.  

On the other hand, ground-based observations are less immune to single-point failure such as a failed launch!  There is also the benefit of being flexible in allowing to improve the instrumentation if needed.  Another benefit is the possibility of having very long observations (decades) usually not possible for space missions due to the cost penalty of equipment lifetime.  For instance, a mission such as SoHO that has been running for the past 17 years was designed for a 2-year lifetime.  Designing a mission of a 17-year mission right for the beginning would have required the spacecraft contractor to plan for 20-year lifetime of the platform, associated with a very high cost.

The benefit of space mission really lies in the scientific needs and the so called "need to go to space".  Space missions can only be justified on the real gain compared to what cannot be done from the ground.  Such a need to go space required hard work at the time of the design of the PRISMA\footnote{Probing Rotation and Interiors of Stars: Microvariability and Activity} mission study \citep{PRISMA91,PRISMA93}.  In that latter case, the scintillation noise was the key driver for going to space combined with the absence of interruption (as a collateral benefit).

\section{Historical perspective of detection}
\subsection{Ground based efforts}
The search for stellar like oscillations on bright stars started in the mid 1980's using resonance spectrometer \citep[Procyon, $\alpha$ Cen A, by ][]{Gelly1986}, using equivalent linewidth \citep[$\beta$ Hyi by][]{Frandsen1987}, using a Fabry-Perot \citep{Ando1988}, with a spectrophotometer \citep[$\alpha$ Cen A by][]{Brown1990}, with a photometer \citep{Belmonte1990, Belmonte1990a,Belmonte1990b}.  

Several additional attempts were made at detecting the modes on Procyon using spectrophotometers: the first credible detection of the mode envelope by \citet{Brown1991}, several other claimed detection \citep{Innis1991, Bedford1993,Mosser1998}, the first measurement of the large separation by \citet{Martic1999}, the first measurement of individual modes and the departure from the regular mode spacing by \citet{Martic2004} using a 2-station network, confirmation of the previous measurements by \citet{Eggenberger2004,Claudi2005,Leccia2007,Mosser2008}.  Finally, a coordinated campaign, including 8 sites distributed around the globe and 11 different instruments, provided the definite answer about the p modes in Procyon \citep{Arentoft2008}.  The campaign provided individual mode frequencies, mode lifetime and a clear echelle diagramme \citep{Bedding2010}.  The noise achieved by that campaign could be compared with the requirement we set above for the noise level in stellar radial velocity.  \citet{Arentoft2008} provided a value of 20 (cm/s)$^{2}\mu{\rm Hz}^{-1}$ at 1000 $\mu$Hz, and of 4 (cm/s)$^{2}\mu{\rm Hz}^{-1}$ above 2000 $\mu$Hz (flat) which are about two to three times lower than the values required by Fig.~\ref{noise_level}.

Detection of modes on $\eta$ Boo were also claimed by \citet{HKTB1995} using equivalent linewidth measurement which could not be confirmed in stellar radial velocity by \citet{Brown1997}.  Nevertheless, \citet{HK2003} confirmed their earlier finding using equivalent linewidth in combination with stellar radial velocity.  The levels quoted in \citet{HK2003} are consistent with a signal-to-noise ratio of 1 when compared with Fig.~\ref{noise_level}.  The undisputed detection of modes in $\eta$ Boo came from \citet{Carrier2005} who reported several modes using a 2-station network at a signal-to-noise level higher than 3; the background noise was about 270 (cm/s)$^{2}\mu{\rm Hz}^{-1}$ at 750 $\mu$Hz about 2.7 higher than the level required by Fig.~\ref{noise_level}.

$\alpha$ Cen A was also extensively studied by \citet{Brown1990, Pottasch1992, Edmonds1995} using stellar radial velocity measurements.  They all reported noise level above 160 (cm/s)$^{2}\mu{\rm Hz}^{-1}$ which is well above the required noise level of Fig.~\ref{noise_level} at 2400 $\mu$Hz about 20 times lower, i.e. with an effective signal-to-noise ratio of 0.5.  From this point of view, \citet{Brown1990} and \citet{Edmonds1995} were coherent in providing only upper limit to the amplitude of the p modes whereas \citet{Pottasch1992} reported detection.  Finally the detection of p modes in $\alpha$ Cen A was confirmed by \citet{Bouchy2001} with a noise level of 8 (cm/s)$^{2}\mu{\rm Hz}^{-1}$ commensurate with the required value of Fig.~\ref{noise_level}.  This work was the real start of asteroseismology since it provided a clear and indisputable detection of p modes in another star than the Sun.

\citet{Frandsen2002} reported the detection of solar-like oscillation in a red giant.  This was also the start of a renewed interest in red giants.  Detection of solar-like modes was also reported in a few other stars using stellar radial velocities \citep[See][for details]{Bedding2011}.

In comparison, the detection of p modes in intensity was not so successful.  There were several attempts by \citet {Belmonte1990, Belmonte1990a,Belmonte1990b} which none could be confirmed since the amount of noise was above 1700\,ppm$^{2}\mu{\rm Hz}^{-1}$, a level at least 100 times larger than the highest values of Fig.~\ref{noise_level}.  The use of Charge Coupled Devices (CCD) was introduced by \citet{Gilliland1988} showing that many stars could be observed; they provided a noise level of about 87\,ppm$^{2}\mu{\rm Hz}^{-1}$ lower than the previous figure but still too high for a credible mode detection.  They improved the technique by using larger telescope area for reaching a value of 25\,ppm$^{2}\mu{\rm Hz}^{-1}$ at 1000 $\mu$Hz \citep{Gilliland1993}, resulting in a typical signal-to-noise ratio of 1 according to Fig.~\ref{noise_level}.  Nevertheless, if the detection of modes in solar-like stars is difficult in intensity from the ground, the technique can be easily applied to red giants having their frequency of maximum power below 200 $\mu$Hz, implying that the mode amplitudes are also much larger \citep{Stello2007}.  Obviously the detection of p modes in intensity would be far easier from space due to the absence of scintillation.

\subsection{Space-based efforts}
As we outlined before, the combination of multi-objects observation in intensity combined with space observations was an attractive perspective for doing asteroseismology on thousands of stars.  The design of such a space mission started in 1980's with the submission of the EVRIS\footnote{Etude de la Variabilit\'e, de la Rotation et des Int\'erieurs Stellaires} mission to the French space agency CNES\footnote{Centre National d'Etudes Spatiales} \citep{Baglin1993}.  It eventually lead to a series of study performed at ESA\footnote{European Space Agency} for various missions such as PRISMA, STARS and {\it Eddington} \citep{PRISMA93,Fridlund1995,Favata2004}.  A full account of the history of space missions can be found in \citet{Roxburgh2002} and references therein.  Table~\ref{comparison} provides a comparison of the various asteroseismic missions planned, launched, being studied or to be launched.  

\begin{table}[!]
\caption{Main characteristics of past, present and future space missions enabling asteroseismology.  The first column is the name of the project.  The second column is the status at which it was stopped or the launch date.  The third column is the effective diameter of the telescope.  The fourth column is the field of view.  The fifth column is the limiting magnitude at which the number of stars of the sixth column are given for a photometric performance provided by the seventh column.}
\centering
\begin{tabular}{c c c c c c c} 
\hline
Project&Status&$D$&FOV &$m_{\rm V}$&Number&Noise\\
&or Launch&(in cm)& (in deg $\times$ deg)&&of stars&(in ppm$^{2}\mu{\rm Hz}^{-1}$)\\
\hline
PRISMA&Phase A&40& 1.5 $\times$ 1.5& $<$8& 2000 & \\
STARS&Phase A&80&1 $\times$ 1&$<$8&2500&\\
{\it Eddington}&Phase B&120& 5 $\times$ 5& $<$11&2000&6\\
MOST&2003&15&0.4 $\times$ 0.4&$<$6& $<$6 & 5.7\\
CoROT&2006&25&1 $\times$ 1&$<$7&10&1.7\\
{\it Kepler}&2009&95&10.5 $\times$ 10.5&$<$12&1300&17.6\\
PLATO&Phase B$\dagger$&67&47 $\times$ 47& $<$11&85000&4.2\\
TESS&2017&10&23 $\times$ 90&$<$12&5$\times$10$^5$&7.6\\
\hline
\end{tabular}
$\dagger$ Still ongoing in 2013 since it is in competition for an M3 mission at ESA.
\label{comparison}
\end{table}

None of the three previous ESA studies became missions even if {\it Eddington} came very close to such a status.  The precursor to all asteroseismic mission EVRIS was launched in November 1996 aboard the MARS96 mission which eventually failed to be put in orbit.  This event put forward on the scene, the CoROT mission which had been submitted to CNES back in 1993 \citep{Catala1995,Roxburgh2002}.  After some momentary interruptions due to governmental intrusion, the CoROT mission was finally launched in December 2006.  

The CoROT mission was preceded by a simple mission akin to the EVRIS mission: the MOST\footnote{Microvariability and Oscillations of Stars} mission of the Canadian Space Agency, launched in June 2003 \citep{Matthews2000}.  The report by \citet{Matthews2004} of the non-detection of p modes in Procyon by MOST started an interesting controversy.  The photometric levels observed at 1000 $\mu$Hz was about 30\,ppm$^{2}\mu{\rm Hz}^{-1}$ which is a factor 10 higher than the required level by Fig.~\ref{noise_level}, i.e. a potential signal-to-noise ratio of 1.  This is this high level of noise (due to Earth scattered light) which was the responsible for the lack of p-mode detection in Procyon as shown by \citet{Bedding2005}.

A special mention to the WIRE\footnote{Wide Field Infrared Explorer} mission which failure was the {\it real} and unplanned start of asteroseismology from space.  WIRE was launched in March 1999 and lost all of its cryogenic capabilities shortly after that.  \citet{Buzasi2000} then saw the opportunity of doing asteroseismology using the CCD of the 5-cm diameter star tracker of the spacecraft;  Tens of stars were measured by the star tracker \citep{Buzasi2004}.  Using WIRE data, \citet{Schou2001} claimed the detection of modes in $\alpha$ Cen A without providing a properly normalised power spectrum, hence making difficult any comparison with theoretical expectations.  \citet{Bruntt2005} provided a value of 8\,ppm$^{2}\mu{\rm Hz}^{-1}$ at 1000 $\mu$Hz for Procyon making their detection of the hump of mode power credible; the required value as given by Fig.~\ref{noise_level} is about three times as low.

Asteroseismic space mission finally came to fruit with the launch of CoROT in December 2006 and of {\it Kepler} in March 2009.  The next section is devoted to their contribution to this two-stage revolution.

\section{The revolution}
\label{revolution}
\subsection{The CoROT mission}
The first revolution was indeed French with the launch of the CoROT mission in 2006.  To date there has been 666 papers (no pun intended) published since January 2007 bearing CoROT in their title.  The highlights of the main asteroseismic discovery of CoROT are available in \citet{Baglin2012} such as the so-called F-star problem \citep{Appourchaux2008}, the detection of mixed modes \citep{Deheuvels2010} and the study of red giants  \citep{DeRidder2009}.  

\begin{figure}
\center{
\hbox{
\includegraphics[width=0.35\textwidth,angle=90]{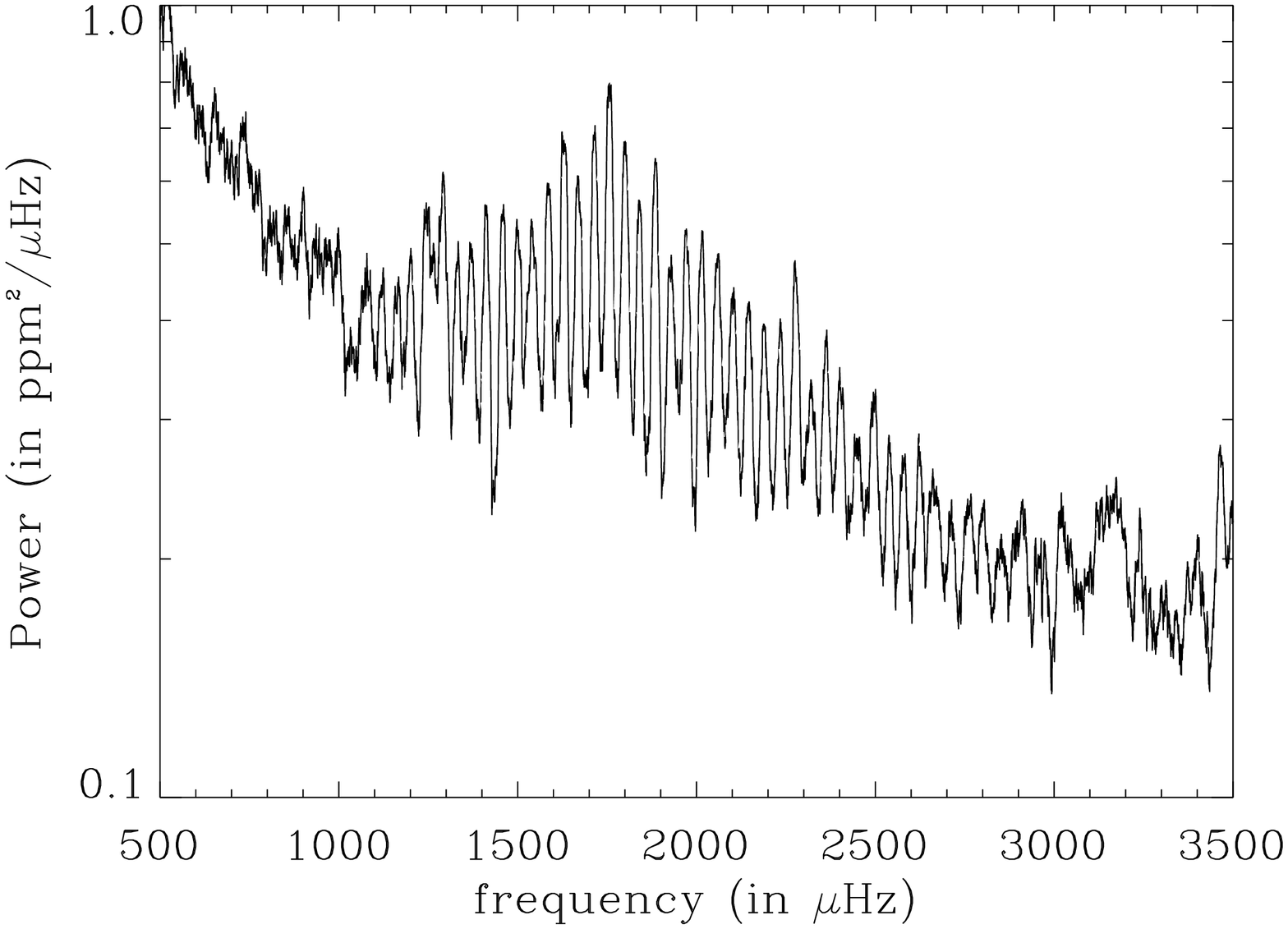}
\hspace{0.2 cm}\includegraphics[width=0.44\textwidth,angle=0]{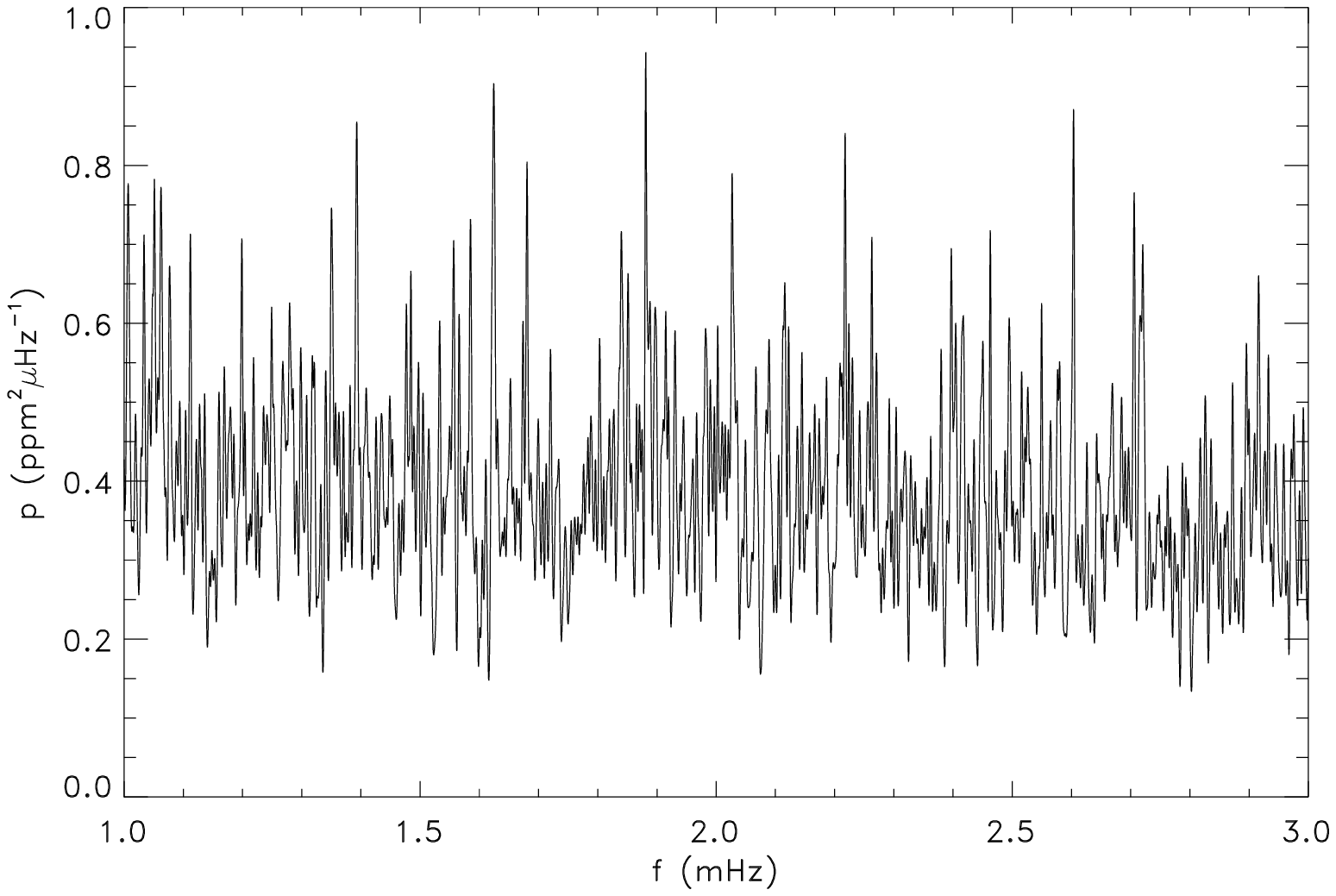}
}
}
\caption[CoROT power spectra]{Power spectra as a function of frequency for HD49933 \citep{Appourchaux2008} (left) and for HD175726 \citep{Mosser2009}.}
\label{Corot_ps}
\end{figure}

The first clear detection of p modes by CoROT was reported by \citet{Appourchaux2008} who provided a noise level of 0.3\,ppm$^{2}\mu{\rm Hz}^{-1}$ at 1800 $\mu$Hz and a typical signal-to-noise ratio of 10 for the highest amplitude.  This results is not in agreement with Fig.~\ref{noise_level} because with such an amplitude the modes should have had a signal-to-noise ratio of 1.  The reason for the discrepancy is the fact that the assumed linewidth in Fig.~\ref{noise_level} was 8 times lower than measured by \citet{Appourchaux2008}, hence with such a measured linewidth the {\it theoretical level} should have been 8 times lower thereby recovering a signal-to-noise ratio close to 10.  The noise level encountered in other solar-like stars ranged from 0.3 to 1\,ppm$^{2}\mu{\rm Hz}^{-1}$ at 1800 $\mu$Hz depending on the magnitude of the star \citep{Michel2008}; these levels are all lower than the specification of Fig.~\ref{noise_level}.  Figure~\ref{Corot_ps} shows the comparison of noise levels for two CoROT stars for which modes were detected or not.

The detection of solar-like oscillations in hundred of red giants is also a landmark of the CoROT legacy \citet{DeRidder2009}.  The noise levels as observed by \citet{Hekker2010} for several red giants at around 25 $\mu$Hz range from 1000 to 5000\,ppm$^{2}\mu{\rm Hz}^{-1}$.  It is consistent with a scaling by $\nu_{\rm max}^2$ of Fig.~\ref{noise_level}.   

\subsection{The {\it Kepler} mission}
\begin{figure}
\center{
\vbox{
\includegraphics[height=10cm,angle=0]{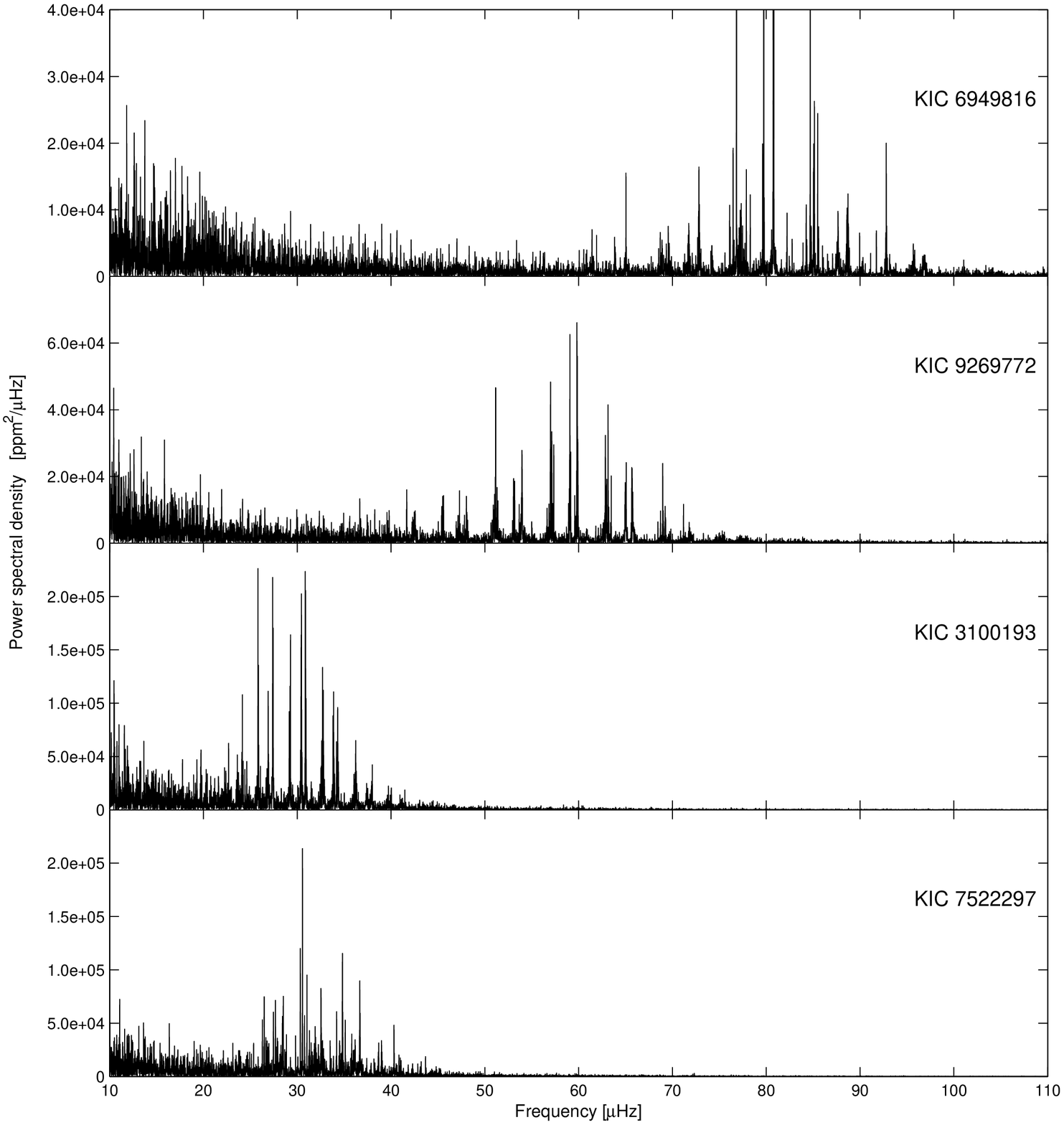}
\includegraphics[height=10.7cm,angle=0]{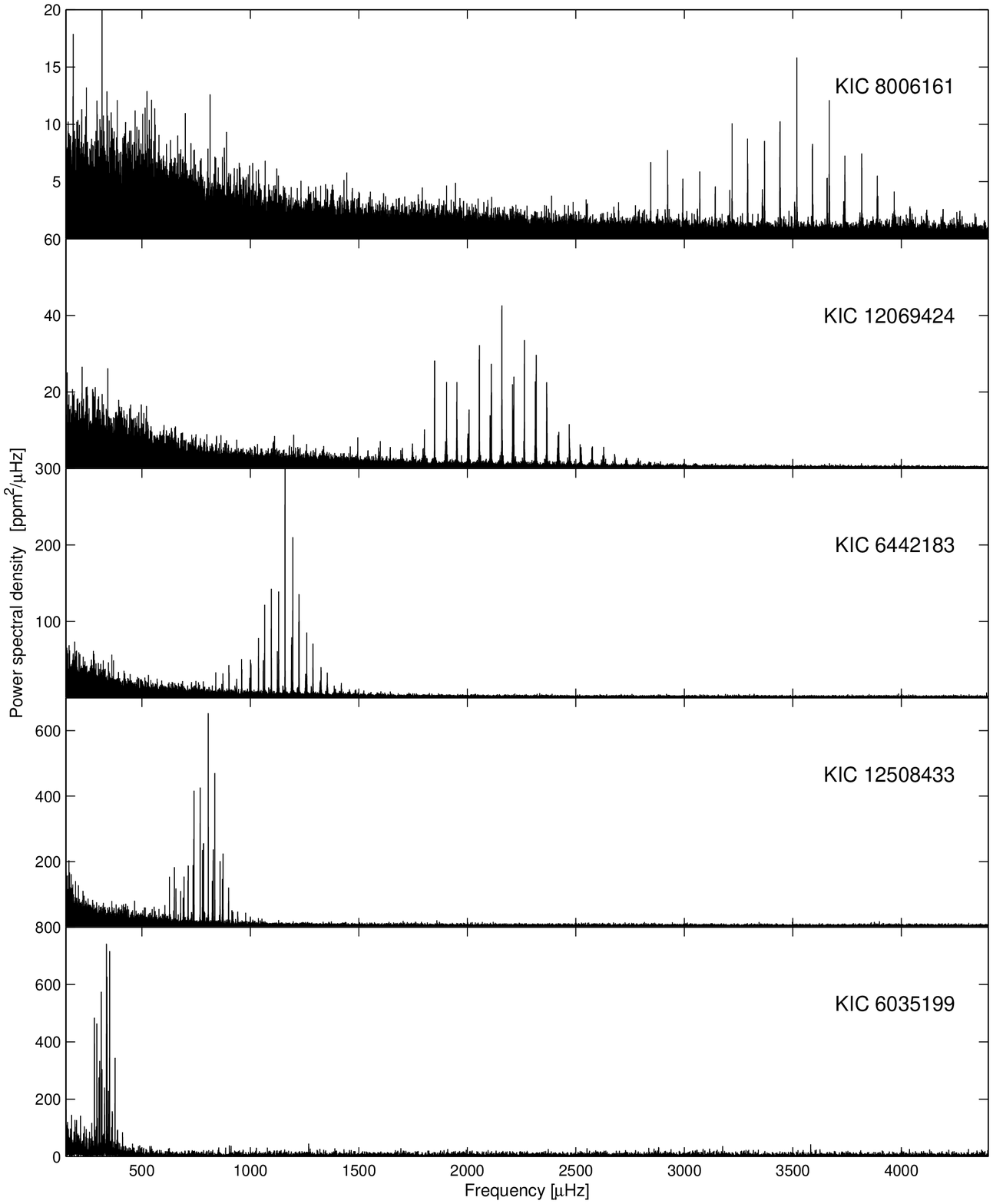}
}
}
\caption[{\it Kepler} power spectra]{Power spectra as a function of frequency for red giants (left and solar-like stars (right) \citep[Stolen from][]{WJCAM2013}.}
\label{Kepler_ps}
\end{figure}

\begin{figure}
\center{
\includegraphics[height=7cm,angle=0]{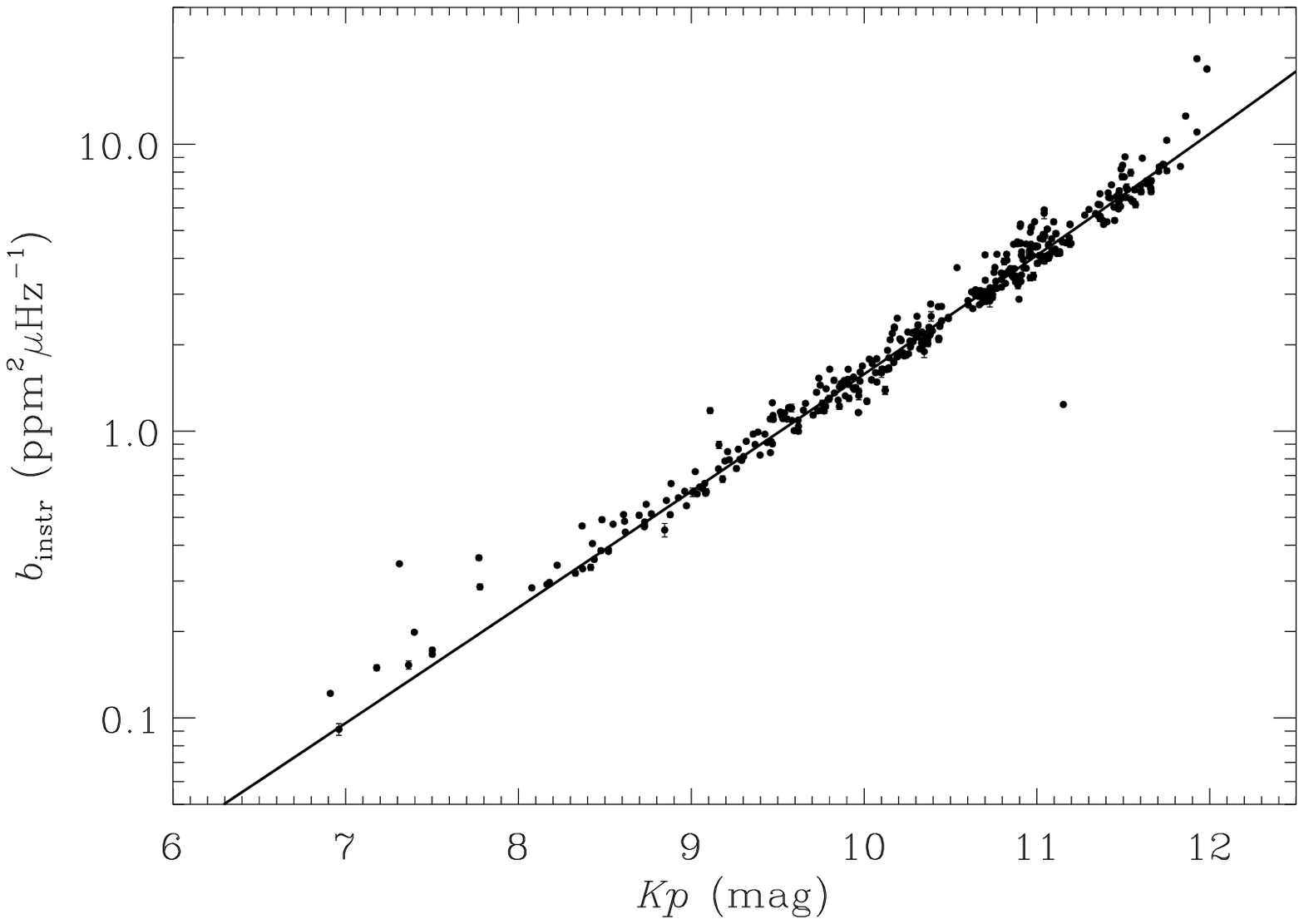}
}
\caption[{\it Kepler} instrumental noise]{{\it Kepler} instrumental noise as a function of magnitude \citep[Stolen from][]{Chaplin2011a}.}
\label{Kepler_inst}
\end{figure}

The second revolution was American with the launch of the {\it Kepler} mission in 2009.  To date there has been more than 1200 papers published since March 2009 bearing {\it Kepler} in their title.  There has been ample papers describing the detection of solar-like p modes across the Hertzprung-Russell diagram with {\it Kepler} but the reader can refer to \citet{WJCAM2013} for a review of the main findings.  Figure~\ref{Kepler_ps} gives a comparison of the power spectra obtained for various stars obtained with the {\it Kepler} mission showing clearly p modes being detected.  The noise as observed by {\it Kepler} is limited by photon noise as shown by \citet{Chaplin2011a} and reproduced in Fig.~\ref{Kepler_inst}.  In comparison, CoROT has a higher instrumental noise mainly due to the its effective telescope diameter being 3.8 smaller.  The noise level of 0.15\,ppm$^{2}\mu{\rm Hz}^{-1}$ obtained for a 5.77 magnitude star \citep{Appourchaux2008} would have been obtained for a magnitude of 8.7 with the theoretical {\it Kepler} telescope diameter.  For the true {\it Kepler} telescope, this previous noise figure is only reached for star brighter than 7.5 magnitude.  Nevertheless, it is now clear than for any star brighter than the 11$^{\rm th}$, the {\it Kepler} performance are better than the specifications of Fig.~\ref{noise_level}.

\section{How to push the envelope?}
With about 30 years of development, {\it modern} asteroseismology has come a long way from the age of non-detection to the age of revolution.
The question the reader may ask is the following: what is left to do that would provide viable Science return?  

History taught us that many unexpected
results were obtained regarding the detection of solar-like oscillation modes in red giants.  From this point of view, there is a renewed interest in red giants and sub giants because these stars exhibit mixed modes with high amplitude and long lifetime sampling the very core of these stars.  The sounding of these stellar cores already provided very useful information about the internal rotation of these stars \citep{Deheuvels2012,Mosser2012}.

Beyond the use of field stars that are useful for doing ensemble asteroseismology \citep{Chaplin2011}, binary stars are the next targets because these stars have a common age, chemical composition and distance.  Only a few of such binary systems have been studied such as $\alpha$ Cen A and B \citep{FB2002,FB2003}; Cyg A and B \citep{Metcalfe2012}; Luke and Leia ({\it Kepler} binary detected by T.White, private communication); and Krazy and Ignatz ({\it Kepler} binary detected by the first author of this paper); seismic red giant (Gaulme and Appourchaux, private communication).  According to William Chaplin and Andrea Miglio (private communication), we can expect about 1\% of binaries showing both a detectable seismic signals.  The prospects for detecting such binaries in solar-like stars, sub-giants and red giants would call for gigantic survey of the sky in order to have about thousands detectable seismic binaries.  

 Another step in going further would be to do asteroseismology for stars in open clusters already anticipated with the PRISMA mission \citep[][and references therein]{PRISMA93}. This provides the opportunity to study stars with a common age, chemical composition and distance -- but with a range in masses and evolutionary stages.   In this respect, we note that many clusters have detached eclipsing binaries which further provide mass and radius fix-points along the cluster sequence with a 1\% accuracy.  In combination with the stellar distances from the GAIA\footnote{Global Astrometric Interferometer for Astrophysics} mission and  asteroseismic measurements this would yield very strong constraints  on stellar models.  At this time of writing, the main work on cluster seismology is that from the {\it Kepler} satelite, see for example \citet{Stello2011} and \citet{Corsaro2012} and references therein.  There are four open clusters in the {\it Kepler} field of view in which NGC 6811, 6819 and 6791 contain at least one detached eclipsing binary \citep{Leitner2013,Jeffries2013,Brogaard2012}. Oscillations are measured for the red-giant branch stars in these clusters.  It would seem that a slightly scaled down {\it Kepler}-like mission with a diameter of 60 cm with a smaller field of view, smaller pixel size (for the crowded cluster fields) would be an ideal cluster mission. For the nearest open clusters this would provide asteroseismology from the main-sequence  to the giants in several open clusters and would even allow the red-giant branch stars in the closest globular clusters to be studied.  The PLATO\footnote{PLAnetary Transits and Oscillations of stars} Mission in competition for selection for an M3 slot for the Cosmic Vision programme of ESA will be able to detect solar-like oscillations in solar-like stars in several open clusters \citep{Catala2009}.  This will not be achieved by the recently selected NASA mission TESS\footnote{Transiting Exoplanet Survey Satellite}.

With the end of the CoROT mission due to the failure of the data acquisition system providing no stellar images, and the recent news that the {\it Kepler} mission is now using a special pointing mode for compensating the loss of one of its reaction wheel, the prospects of having high quality photometric data for asteroseismology might be somewhat reduced.  Nevertheless, a call for White Paper for using the new pointing mode of {\it Kepler} resulted in the possibility of looking at open clusters not in the original Cygnus field but in the ecliptic \citep{Chaplin2013}.  If this proves not feasible then the prospects of doing asteroseismology from space are now resting on the shoulders of the PLATO mission.  With a potential launch in 2024, there is now a mere 10 years to wait for fresh data, a gap that could be dreadful for anyone (such as the first author of this paper) having started to work on studying such a mission back at the end of the 90's !  Fortunately, the younger generations that grew up with CoROT and {\it Kepler} might be readier to wait during that decade.  The coming decade could be a good opportunity to re-measure the classical stellar parameters that are needed for stellar modelling using asteroseismic data \citep{Apogee2013}.

From these prospects, a project such as SONG (song.au.dk) is worth pursuing.  SONG is a network of 8 nodes located at existing observatories. There will be four nodes in each hemisphere equipped with identical telescopes and instrumentation.  At each node, the instruments will be an imaging camera and a high-resolution spectrograph (resolution of 10$^5$) . SONG will fill a gap in the quality of the detected modes compared to what is currently being achieved in intensity.  Furthermore by directing attention to the nearest and brightest stars it will be possible to obtain highly precise angular diameters, distances and temperatures from astrometry (Hipparcos, GAIA) and interferometry. Such external constraints are much needed for testing and evaluating the quality of stellar models.

Finally, let us dream on the instrument of the future: a hyper-telescope able to image the stars for imaging higher degree modes than currently possible (say $l > 50$).  With that capability, we will be able to study and understand the internal rotation of stars as a function of mass and age.  This will be a great asset for improving our understanding of stellar physics.  An hyper-telescope, likely to be a space mission, would image the star using a set of small telescopes distributed over an equivalent diameter of the order of 1 km.  The focal plane instrument could either measure stellar radial velocity using a Mach-Zehnder type of instrument such as the {\it Echoes} instrument developed for doing Jovian seismology aboard the ESA's JUICE mission \citep{Soulat2012}; or measure intensity using a CCD array.  Such a mission called the {\it Stellar Imager} was proposed for the US decadal survey in 2010 by \citep{Carpenter2010}.  This kind of mission requires that the small telescopes, at a few kms distance from each other,  interferometrically combine their optical beams at the focal plane.  This challenging mission has been studied by \citep{Labeyrie2009} and a road map has even been laid out \citep{Labeyrie2008}.  Time will tell us whether this mission is the ultimate asteroseismic mission.


\bibliographystyle{aa}
\bibliography{thierrya}






  \cleardoublepage





\end{document}